\documentclass[12pt]{article}

\renewcommand{\baselinestretch}{1.4}
\usepackage{amsmath}
\usepackage{graphicx}
\usepackage{enumerate}
\usepackage{natbib}
\usepackage{amsfonts}

\newcommand{\cN}{{\mathcal{N}}}

\newcommand{\RR}{\mathbb{R}}

\newcommand{\0}{{\mathbf{0}}}

\renewcommand{\b}{{\mathbf{b}}}

\newcommand{\s}{{\mathbf{s}}}
\renewcommand{\t}{{\mathbf{t}}}

\newcommand{\w}{{\mathbf{w}}}
\newcommand{\x}{{\mathbf{x}}}

\newcommand{\E}{{\mathbf{E}}}
\renewcommand{\P}{{\mathbf{P}}}

\renewcommand{\S}{{\mathbf{S}}}

\newcommand{\V}{{\mathbf{V}}}
\newcommand{\W}{{\mathbf{W}}}
\newcommand{\X}{{\mathbf{X}}}
\newcommand{\Y}{{\mathbf{Y}}}
\newcommand{\Z}{{\mathbf{Z}}}

\newcommand{\hbeta}{{\hat{\beta}}}

\newcommand{\bbeta}{\boldsymbol{\beta}}
\newcommand{\bsigma}{\boldsymbol{\sigma}}
\newcommand{\hbbeta}{\boldsymbol{\hat{\beta}}}
\newcommand{\hbsigma}{\boldsymbol{\hat{\sigma}}}

\newcommand{\btheta}{\boldsymbol{\theta}}
\newtheorem{theorem}{Theorem}[section]
\newcommand{\vX}{\vec{\X}}
\newcommand{\vV}{\vec{\V}}

\begin{document}

\title{Calibrated Percentile Double Bootstrap \\ For Robust Linear Regression Inference}

  \author{Daniel McCarthy
\hspace{.2cm} \\
    Department of Statistics, University of Pennsylvania \\
    and \\
    Kai Zhang\thanks{
    Lawrence Brown and Daniel McCarthy were partially supported by NSF grant DMS-1512084. Kai Zhang was partially supported by NSF grant DMS-1309619, and is also partially supported by the NSF under Grant DMS-1127914 to the Statistical and Applied Mathematical Sciences Institute. The content is solely the responsibility of the authors and does not necessarily represent the official views of the National Science Foundation.} \\
    Department of Statistics, University of North Carolina at Chapel Hill \\
\\
Lawrence Brown, Richard Berk,  Andreas Buja, Edward George and Linda Zhao\\
Department of Statistics, University of Pennsylvania}
  \maketitle

\begin{abstract}
We consider inference for the parameters of a linear model when the covariates are random and the relationship between response and covariates is possibly non-linear. Conventional inference methods such as {\tt z} intervals perform poorly in these cases. We propose a double bootstrap-based calibrated percentile method, {\tt perc-cal}, as a general-purpose CI method which performs very well relative to alternative methods in challenging situations such as these. The superior performance of {\tt perc-cal} is demonstrated by a thorough, full-factorial design synthetic data study as well as a real data example involving the length of criminal sentences. We also provide theoretical justification for the {\tt perc-cal} method under mild conditions. The method is implemented in the {\tt R} package `perccal', available through CRAN and coded primarily in {\tt C++}, to make it easier for practitioners to use.
\end{abstract}

\begin{quote} \small
{\bf Keywords}:  Confidence intervals; Edgeworth expansion; Second-order correctness; Resampling
\end{quote} \normalsize

\newpage
\setcounter{page}{1}
\vspace{-0.15in}
\section{Introduction}  \label{intro}

In many applied settings, practitioners would like to make interpretable statements such as the expected average difference in a response variable, $Y$, associated with a unit difference in a covariate of interest, $X_j$, controlling for all other predictors. In situations like these, practitioners often run linear regressions despite the fact that the true but unobservable relationship between $Y$ and $X_j$'s may be non-linear. Doing so may be sensible when the utility of being able to make more interpretable statements such as the one above outweighs the cost of possible model bias, which may hard to discern (particularly in multivariate settings). An important challenge is how practitioners can produce valid inference upon their estimates of the true population-level best linear approximation for the relationship between predictor and response in these settings. We denote the target of interest by $\bbeta$ (for more detail, please see Section \ref{sec:framework}). Our aim with this paper is to help practitioners perform better inference for $\bbeta$ in these situations.

\cite{bujamodels} call much-needed attention to this issue, showing that when the relationship between response $Y$ and covariates $\vX=(1,X_1,\ldots,X_p)^T$ is truly non-linear with noise that is possibly heteroskedastic, and when $\vX$ is itself random, standard linear model theory standard errors are asymptotically invalid. They show that the ``sandwich estimator'' of standard error does provide asymptotically correct inference for the population slopes, even when non-linearity and heteroskedasticity are present, and $\vX$ is random.

While the sandwich estimator may provide asymptotically valid inference, practitioners will also be understandably interested in better understanding how the finite sample performance of various methods of inference compare, as well as the asymptotic properties of those methods. We will show that in our setting, empirical coverage of population regression slopes deteriorates considerably for all traditional confidence interval methods. The primary contribution of this paper is to shine new light on these issues, proposing and studying an inference method which is convincingly superior to the sandwich estimator, and making a very fast implementation of this proposed method accessible to practitioners.

We propose a double bootstrap-based calibrated percentile method, {\tt perc-cal}. The seminal work of Peter Hall shows the advantages of double bootstrap approach in classical settings involving  population means. Population slopes as defined here are a more complex, non-linear object. For example, \cite{hall1992bootstrap} studies univariate data without model misspecification. The methods he uses, then, need to be augmented with additional material about Edgeworth expansions that is adapted from \cite{jensen1989}. For the first time, we prove in Section \ref{asympt-th} and in the appendix that even when $Y$ and $\vX$ have a non-linear joint distribution, and $\vX$ is random, that under relatively mild regularity conditions the rate of coverage error of {\tt perc-cal} for two-sided confidence intervals of the best linear population slopes between a response variable $Y$ and $p$-dimensional covariates $\vX$ is $\mathcal{O}(n^{-2})$. In contrast, conventional methods achieve a rate of coverage error of $\mathcal{O}(n^{-1})$. We then show in a Monte Carlo study that {\tt perc-cal} performs better than traditional confidence interval methods, including the BCa method (\cite{efron1987better}), and other Sandwich-based estimators discussed in \cite{cribari2007inference} and \cite{mackinnon2013thirty}. Our study is similar in structure to the simulation study that was performed in \cite{gonccalves2005bootstrap}, but modified to study a very wide variety of misspecified mean functions. We follow up this synthetic simulation study with a real data example involving a criminal sentencing dataset, and show that {\tt perc-cal} once again performs satisfactorily. We have released an {\tt R} package, `perccal' (\cite{mccarthy2016perccal}), available through CRAN and coded primarily in {\tt C++}, so that practitioners may benefit from a fast implementation of {\tt perc-cal} for their own analyses.

We argue the combination of theoretical and empirical justification presented in this paper supports the claim that {\tt perc-cal} is a reliable confidence interval methodology that performs well in general, even in the presence of relatively severe model misspecification. The remainder of the paper is organized as follows. Section \ref{lit-rev} provides a review of confidence interval methods. Section \ref{asympt-th} presents the theoretical results. Section \ref{numer-studies} compares the performance of {\tt perc-cal} with that of other often more commonly used confidence interval estimators in synthetic and real data settings. Section \ref{disc-concl} provides concluding remarks, and an Appendix gives all of the proofs.

\section{Literature Review}  \label{lit-rev}
\subsection{Review of Bootstrap Confidence Intervals}  \label{review-CI-methods}
There is a very wide variety of bootstrap methods that have been proposed in the literature to compute $(1-\alpha)$ confidence intervals. These methods include Efron's percentile method (\cite{efron1981nonparametric}, page 146), Hall's percentile approach (\cite{hall1992bootstrap}, page 7), and Hall's percentile-$t$ method (\cite{hall1988theoretical}, page 937). Other forms of bootstrap CIs include symmetric CIs (\cite{hall1992bootstrap}, page 108) and short bootstrap confidence intervals (\cite{hall1992bootstrap}, page 114). In general, performance of these methods depends upon the properties of the data generating process and/or the sample size. We are primarily interested in confidence interval methods that assume much less about the true underlying data generating process, which is usually unknown and often not well behaved in real data applications, making these methods less relevant to the work which follows.

Hall advocates the use of pivotal bootstrap statistics because pivotal bootstrap statistics have higher asymptotic accuracy when the limiting distributions are indeed pivotal (\cite{hall1992bootstrap}, page 83). We emphasize that Hall's preference for pivotal bootstrap statistics, and much of the discussion regarding the relative merits of various confidence interval methods, are based on the {\it asymptotic} properties of these methods. When the sample size is small, these asymptotic considerations do not necessarily reflect the empirical performance of these methods. For example, Hall cautioned that ``our criticism of the percentile method and our preference for percentile-$t$ lose much of their force when a stable estimate of $\sigma^2$ is not available'' (\cite{hall1988theoretical} page 930). Simulation studies that reinforce this include \cite{scholz2007bootstrap}.

Another class of confidence intervals may be formed by replacing the standard error estimator in the standard {\tt z} or {\tt t} interval with a so-called `Sandwich' estimator (\cite{white1980heteroskedasticity}), or one of the many extensions of the Sandwich estimator (\cite{cribari2007inference}). A comprehensive review of Sandwich estimators can be found in \cite{mackinnon2013thirty}, and we will compare these methods with our proposed method in Section~\ref{numer-studies}.

\subsection{Review of Iterative Bootstrap Confidence Intervals}\label{sec:doubleb}
The idea of the iterative bootstrap (or double-bootstrap) was first introduced in \cite{efron1983estimating}. The improvement on coverage probability of CIs was first analyzed in \cite{hall1986bootstrap} and later discussed in more detail in \cite{loh1987calibrating}, \cite{hall1988bootstrap}, \cite{hall1989better}, \cite{martin1990bootstrap}, and \cite{martin1992double}. A comprehensive review can be found in Section~3.11 in \cite{hall1992bootstrap} (see also \cite{efron1994introduction}, page 268). In general, the iterative bootstrap provides more accurate coverage probability at the cost of more computing.

\begin{figure}[ht!]
\centering
\includegraphics[width=0.98\textwidth]{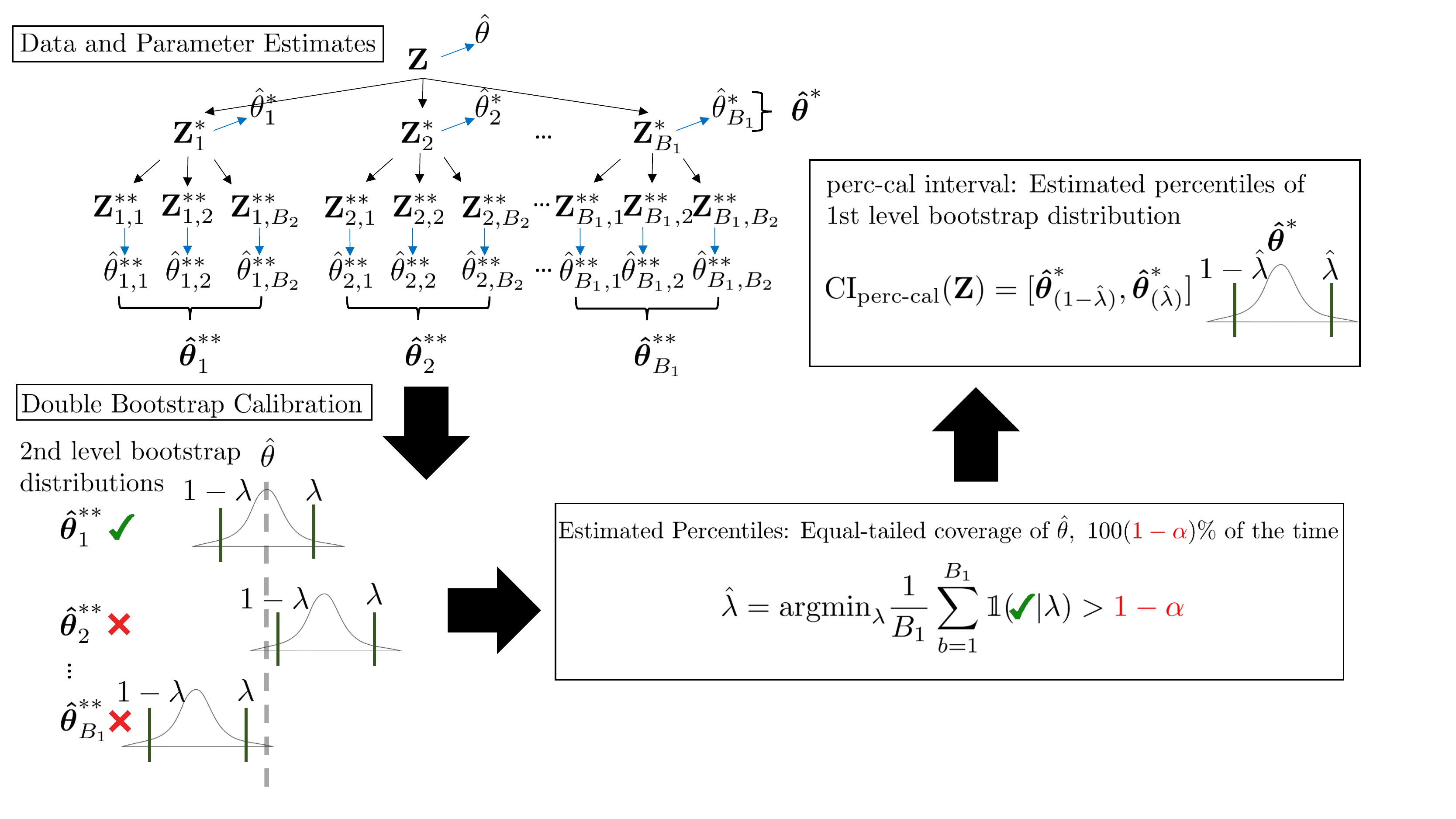}
\caption{{\tt perc-cal} diagram}
\label{perccal-diagram}
\end{figure}

To fix ideas, in this section we shall introduce the proposed double-bootstrap confidence interval method in a univariate case with generic notations. We will extend this procedure to the regression setting in Section~\ref{sec:framework}. We assume that we observe $Z_1,\ldots,Z_m \stackrel{iid}{\sim} F$ for some distribution $F$. Let $\theta=\theta(F)$ be a parameter of our interest. We will estimate $\theta$ through the empirical distribution $\hat{F}(z)={1 \over m}\sum_{i=1}^m I(Z_i \le z).$  The estimator is denoted by $\hat{\theta} = \theta(\hat{F})=\theta(Z_1,\ldots,Z_m)$. The construction of the confidence interval is illustrated in Figure \ref{perccal-diagram} and is described as follows.

\begin{enumerate}
\item For chosen bootstrap sample size $B_1$, obtain bootstrap samples $(\Z_1^{*}, \ldots, \Z_{B_1}^{*})$. Each $\Z_j^*$ consists of $m$ i.i.d. samples with replacement from $\hat{F}$. For chosen bootstrap sample size $B_2$, obtain double bootstrap samples corresponding to all bootstrap samples, $(\Z_{1,1}^{**}, \ldots, \Z_{1,B_2}^{**}, \Z_{2,1}^{**}, \ldots, \Z_{2,B_2}^{**}, \ldots, \Z_{B_1,1}^{**}, \ldots, \Z_{B_1,B_2}^{**})$ in the same manner as in the first-level bootstrap. Denote the empirical distributions by $\hat{F}_j^*$'s, $j=1,\ldots, B_1$, and $\hat{F}_{j,k}^{**}$'s, $j=1,\ldots, B_1, k=1,\ldots,B_2$, respectively.
\item Obtain parameter estimates corresponding to the observed sample, $\hat{\theta}=\theta(\hat{F})$, all bootstrap samples, $(\hat{\theta}_1^{*}, \ldots, \hat{\theta}_{B_1}^{*})$ with $\hat{\theta}_{j}^{*}=\theta(\hat{F}_j^{*})$ and all double bootstrap samples corresponding to all bootstrap samples, $(\hat{\theta}_{1,1}^{**}, \ldots, \hat{\theta}_{1,B_2}^{**}, \hat{\theta}_{2,1}^{**}, \ldots, \hat{\theta}_{2,B_2}^{**}, \ldots, \hat{\theta}_{B_1,1}^{**}, \ldots, \hat{\theta}_{B_1,B_2}^{**})$ with $\hat{\theta}_{j,k}^{**}=\theta(\hat{F}_{j,k}^{**})$.
\item Form $B_1$ double-bootstrap histograms $\hat{\btheta}_1^{**}, \ldots, \hat{\btheta}_{B_1}^{**}$, where each histogram $\hat{\btheta}_j^{**}$ is comprised of all $B_2$ double bootstrap estimates $(\hat{\theta}^{**}_{j,1}, \ldots, \hat{\theta}^{**}_{j,B_2})$ corresponding to the $j$th bootstrap sample and estimate, $\Z_j^*$ and $\hat{\theta}_j$, respectively, $j \in \{1, 2, \ldots, B_1 \}$.
\item Find the largest $\hat{\lambda}$ such that $1/2<\hat{\lambda}<1$ and that $\hat{\theta}$ lies in the $1-\hat{\lambda}$ percentile and the $\hat{\lambda}$ percentile of the histograms $1-\alpha$ proportion of the time.

\item $\hat{\theta}$ lies between the $(1-\hat{\lambda},\hat{\lambda})$ percentiles of the second-level bootstrap distributions $1-\alpha$ proportion of the time. Therefore our {\tt perc-cal} $(1-\alpha)$ interval for $\theta$ is equal to the $(1-\hat{\lambda},\hat{\lambda})$ percentiles of the first-level bootstrap distribution, $[\hat{\btheta}^*_{(1-\hat{\lambda})}, \hat{\btheta}^*_{(\hat{\lambda})}]$.
\end{enumerate}

For a $(1-\alpha)$ left-sided {\tt perc-cal} confidence interval for $\theta$, the only change in the procedure is in Step 4, where one uses the histograms to find the smallest $\hat{\lambda}$ such that $\hat{\theta}$ lies below the $\hat{\lambda}$ percentile of the histograms $1-\alpha$ percent of the time. In what follows, we shall refer the two-sided {\tt perc-cal} interval as $\mathcal{I}_2=[\hat{\btheta}^*_{(1-\hat{\lambda})},\hat{\btheta}^*_{(\hat{\lambda})}]$ and the one-sided {\tt perc-cal} interval as $\mathcal{I}_1=(\infty,\hat{\btheta}^*_{(\hat{\lambda})}].$

A similar double-bootstrap confidence interval is the double-bootstrap-$t$ method which uses the second-level bootstrap to calibrate the coefficient of the bootstrap standard deviation estimate. In practice, both methods can be applied. Hall commented in his book (\cite{hall1992bootstrap}, page 142) that ``either of the two percentile methods could be used, although the `other percentile method' seems to give better results in simulations, for reasons that are not clear to us.'' Here ``the `other percentile method''' refers to confidence intervals $\mathcal{I}_1$ and $\mathcal{I}_2$. Our simulation studies in Section~\ref{numer-studies} demonstrate the same phenomenon. We have observed through simulation that the performance of double-bootstrap-$t$ can be erratic at times due to the instability which arises by relying upon a statistic which has a double bootstrap estimated standard error in its denominator. Some samples will inevitably have very small or even degenerate bootstrap standard errors, making this statistic very large. This issue is particularly acute when sample sizes are small.

Research on optimizing the trade-off between the number of simulations, $B_1$ and $B_2$, in double-bootstrap and the CI accuracy can be found in \cite{BERAN01091987}, \cite{beran1987refinements}, \cite{BOOTH01061994}, \cite{booth1998allocation}, \cite{Lee1999approximation}, among many others. In particular, \cite{Lee1999approximation} study the asymptotic convergence rate of the coverage probability involving $B_1$ and $B_2$, and suggest an adaptive method to optimize the choice of $B_2$ as a function of $B_1$. They note that $B_1$ should be to set to a larger number (for example, $B_1=1000$), and $B_2$ equal to a lesser value. In all simulations which follow, we set $B_1=B_2=2000$. Since the computation of {\tt perc-cal} is reasonably efficient as discussed in Section \ref{disc-concl}, we do not optimize the number of bootstrap samples further but note that further performance gains for robust linear regression inference are a promising area for future research (in particular, with respect to $B_2$).

\subsection{Review of Bootstrap Applications in Conventional Linear Models}
Bootstrap in linear models is studied in Section~4.3 in \cite{hall1992bootstrap}. Hall refers the fixed design case the ``regression model'' and the random design case the ``correlation model.'' Bootstrap estimation and confidence intervals for the slopes, as well as simultaneous confidence bands, are described.

Since the seminal paper of \cite{freedman1981}, the bootstrap has been widely used in regression models because of its robustness to the sample distributions. A review of bootstrap methods in economics can be found in \cite{mackinnon2006bootstrap}. \cite{gonccalves2005bootstrap} consider bootstrapping the sandwich estimator for the standard error when the observations are dependent and heterogeneous. Bootstrap applications under other types of model misspecifications are recently considered in \cite{kline2012higher} and \cite{spokoiny2014bootstrap}. In this paper, we focus on a different case when observations of $(Y,\vX)$ are i.i.d. but the joint distribution is assumption-lean -- we elaborate upon this further in the next section.

\subsection{The Assumption-Lean Framework and Double-Bootstrap Applications}\label{sec:framework}
Conventional linear models assume $\E[Y|\vX]=\vX \bbeta$ for some $(p+1)$-vector $\bbeta$ as regression coefficients, so that $Y$ depends on $\vX$ only through a linear function. While this is commonly assumed in the bootstrap literature, we may not want to require it when performing inference in real data settings because the true relationship may not be linear. Moreover, as first noted in \cite{white1980heteroskedasticity}, a non-linear relationship between $Y$ and $\vX$ and randomness in $\vX$ can lead to serious bias in the estimation of standard errors. \cite{bujamodels} reviewed this problem, and proposed an ``assumption-lean'' framework for inference in regressions. In this framework, no assumption is made regarding the relationship between $Y$ and $\vX$. The only assumptions are on the existence of certain moments of the joint distribution of $\vV=(X_1,\ldots,X_p,Y)^T$. This consideration makes the model very general and thus widely applicable. Readers are referred to \cite{bujamodels} for more details.

Even though a linear relationship between $\E[Y|\vX]$ and $\vX$ is not assumed in an assumption-lean framework, the slope coefficients that are estimated are always well-defined through a population least-squares consideration: the population least-squares coefficients $\bbeta$ minimize squared error risk over all possible linear combinations of $\vX$:
 \begin{equation}\label{eq:bbeta}
    \bbeta=argmin_{\b} \E\|Y-\b^T \vX\|_2^2 = \E[\vX\vX^T]^{-1} \E[Y \vX].
  \end{equation}
This definition of coefficients $\bbeta$ is meaningful in addition to being well defined under minimal assumptions: $\bbeta$ provides us with the best linear approximation from $\vX$ to $Y$, whether or not $\vX$ and $Y$ are linearly related to one another. This setup allows for situations including random $\vX$, non-Normality, non-linearity and heteroskedasticity and we show later that the proposed {\tt perc-cal} method provides better empirical coverage of the true population least-squares coefficients $\bbeta$ on average over a wide variety of data generating processes, even if those data generating processes involve random $\vX$, non-linearity in $E(Y|\vX)$ and/or heteroskedasticity. In contrast, previous research on the double bootstrap has studied functions that are linear or approximately linear.

To estimate $\bbeta$, denote the i.i.d. observations of $\vV$ by $\vV_1,\ldots,\vV_n$ and denote the $n \times (p+1)$ matrix with rows $\vV_1,\ldots,\vV_n$ by $\V$. Denote the distribution of $\V$ by $G$. The multivariate empirical distribution of $\vV$ is then $\hat{G}(\vV)=\hat{G}(x_1,\ldots,x_p,y)={1 \over n}\sum_{i=1}^n I(X_{1,i} \le x_1,\ldots,X_{p,i} \le x_p,Y_i \le y).$ The least squares estimate for $\bbeta$ defined in \eqref{eq:bbeta} is
  \begin{equation}\label{eq:bbeta.hat}
    \hbbeta = (\X^T\X)^{-1} \X^T \Y,
  \end{equation}
where $\X$ is the $n \times (p+1)$ matrix and $\Y$ is the $n \times 1$ vector containing the i.i.d. observations of $\vX$ and $Y$ respectively. Note that each estimate $\hat{\beta}_j$ of $\beta_j$ can be written as a function of $\hat{G}$, $\hat{\beta}_j = \beta_j (\hat{G}).$

The {\tt perc-cal} confidence intervals for each of the slopes $\beta_j$ in $\bbeta$ are constructed similarly as described in Section~\ref{sec:doubleb}. We use the pairs bootstrap first proposed by \cite{freedman1981} and create $B_1$ i.i.d. bootstrap samples, $\V_k^*$, where each matrix $\V_k^*$ consists $n$ i.i.d. samples with replacement from $\hat{G}$. From these samples, one can create the empirical distribution $\hat{G}_k^*.$  To find the proper calibration of $\mathcal{I}_1$ or $\mathcal{I}_2$ (as defined in Section \ref{sec:doubleb}), we sample $B_2$ i.i.d. pairs bootstraps $\V_{k,h}^{**}$ with empirical distributions $\hat{G}_{k,h}^{**}$. The other steps in the construction are identical to those in Section~\ref{sec:doubleb} with $\theta(\cdot)$ replaced by $\beta_j(\cdot)$ with respective empirical distributions as arguments. We note here that although confidence intervals can be built through empirical process theory \citep{Vaart1998}, the accuracy is usually not as good as the proposed {\tt perc-cal} method, as will be discussed in the next section. While \cite{delaigle2015confidence} studies bootstrap confidence bands under a nonparametric regression setting, we are unaware of any existing general bootstrap theory that yields general results for iterated bootstrap confidence intervals for a non-linear function of expectations of non-linear functions.

\section{Asymptotic Theory}  \label{asympt-th}

In this section, we discuss the theoretical properties of {\tt perc-cal}. The following theorem describes the accuracy on the coverage probability of {\tt perc-cal} confidence intervals.
\begin{theorem}\label{thm:asym}
  Consider $n$ i.i.d. observations of the $(p+1)$-dimensional random vector $\vV=(X_1,\ldots,X_p,Y)^T.$ Denote the vector of $Y$ and the continuous $X_j$'s by $\vV_C$ and that of the discrete $X_j$'s by $\vV_D$. Suppose that
  \begin{enumerate}
    \item\label{cond:non-deg} $\exists \ C_0>0$, such that the minimal eigenvalue of the covariance matrix $Var(\vV)$ is larger than $C_0$.
    \item\label{cond:moment} The moment generating function of the random variable $\|\vV\|_2$ exists for all $t \in \RR$: $\E[e^{t\|\vV\|_2}]< \infty$ for all $t \in \RR$.
    \item\label{cond:dist} Conditional on every single value of $\vV_D$, the joint distribution of $\vV_C$ is absolutely continuous with respect to the Lebesgue measure.
  \end{enumerate}
  Consider the population least squares parameter defined in \eqref{eq:bbeta} whose estimate is the sample least squares defined in \eqref{eq:bbeta.hat}. Then for each $1 \le j \le p$, the $(1-\alpha)$ {\tt perc-cal} CI for $\beta_j$ described in Section~\ref{sec:doubleb} and Section~\ref{sec:framework} have the coverage probabilities
  \begin{eqnarray}
    \P(\bbeta_j \in \mathcal{I}_1) &=& 1-\alpha+O(n^{-1}).\\
    \P(\bbeta_j \in \mathcal{I}_2) &=& 1-\alpha+O(n^{-2}).
  \end{eqnarray}
\end{theorem}

The proof of Theorem~\ref{thm:asym} is in Section \ref{proof-asympt} in the Appendix. It uses techniques of the Edgeworth expansion as in Section 3.11.3 of \cite{hall1992bootstrap}, with a focus on the assumption-lean regression model setting where the dependence between $Y$ and $X_j$'s are not necessarily linear. We are not aware of prior investigation of the performance of the double bootstrap confidence intervals under this situation. Moreover, the results in \cite{martin1990bootstrap} and \cite{hall1992bootstrap} accommodates only $X_j$'s satisfying Cram\`{e}r's condition, which excludes distributions such as the Poisson distribution. Discrete distributions for $X_j$ are often encountered in models involving categorical predictors, and the data example we study in Section \ref{numer-studies} involves such covariates. Through a conditioning argument in \cite{jensen1989}, we are able to show that the same performance is enjoyed by a wider class of mixed discrete and continuous $X_j$'s.

Our results show that the coverage probability is $1-\alpha+O(n^{-2})$. Note that for other construction approaches of confidence intervals such as from the sandwich estimator or from the empirical process theory, the resulting one-sided confidence intervals often have a coverage probability of $1-\alpha +O(n^{-1/2})$, and two-sided ones often have a coverage probability of $1-\alpha+O(n^{-1})$ (see Section 3.5.4 and Section 3.5.5 in \cite{hall1992bootstrap}). Thus, the double bootstrap method provides better coverage.

\section{Numerical Studies}   \label{numer-studies}

In this section, we study the performance of {\tt perc-cal} compared to alternative (often more common) methods for forming confidence intervals, including other double bootstrap methods. We first compare {\tt perc-cal} to these other methods using simulated data under a very wide variety of true data generating processes. We then illustrate our approach in a real data example. We will see that {\tt perc-cal} performs very satisfactorily in general.

\subsection{Synthetic Simulation Study}

\subsubsection{Design: Synthetic Simulation}   \label{scatterplots}

We compare the performance of {\tt perc-cal} with 10 other methods that are commonly used for constructing confidence intervals:
\begin{enumerate}
\item [1.] Standard normal interval: {\tt z} (\cite{efron1994introduction}, pp. 168).
\item [2-6.] Five sandwich variants: {\tt sand1}, {\tt sand2}, {\tt sand3}, {\tt sand4} and {\tt sand5} (\cite{mackinnon2013thirty} provides a review of these methods, denoted there by H1, H2, H3, HC4 and HC5).
\item [7.] Hall's Studentized interval: {\tt stud} (\cite{hall1988theoretical}).
\item [8.] Hall's ``bootstrap-t'' method: {\tt boot-t} (\cite{efron1994introduction}, pp. 160-162).
\item [9.] Efron's BCa interval: {\tt BCa} (\cite{efron1987better}).
\item [10.] Single percentile method: {\tt perc} (\cite{efron1994introduction}, pp. 170).
\end{enumerate}

\bigskip

We consider a very wide range of underlying true data generating models, to obtain a more general understanding for how these confidence interval methods compare against one another in a wide variety of data settings, for large sample sizes as well as small. The data generating models represent a full factorial design of the following 48 factors, after excluding non-denerate combinations (i.e., combinations for which the conditional mean is not finite):
\begin{itemize}
\item Simple regression - one predictor, $Y = \beta_0 + \beta_1 X + \epsilon$
\item Sample size $n$ = 32, 64, 128, 256
\item Relationships between Y and X: (1) $Y = X + e$; (2) $Y = \exp(X) + e$; (3) $Y = X^3 + e$.
\item Distribution of X: (1) $X \sim \mathcal{N}(0,1)$, (2) $X \sim \exp(\mathcal{N}(0,1))$.
\item Noise: $\epsilon \sim (1) N(0,1); (2) \ |X|*\mathcal{N}(0,1); (3) \ \exp(\mathcal{N}(0,1)).$
\end{itemize}

In each of the above cases, we use 2000 first and second-level bootstrap samples for all bootstrap methods ($B_1=B_2=2000$). We obtain empirical coverage figures for the slope coefficient in the regression, $\beta_1$. Results are averaged over 500 replications to reduce the empirical standard error of the resulting intervals to below 1.5\% on average across scenarios and methods. We present the results for a target coverage of 90\%, without loss of generality (results for a target coverage of 95\% are qualitatively the same).

\subsubsection{Results}   \label{scatterplots}
\begin{figure}[ht!]
\centering
\includegraphics[width=0.98\textwidth]{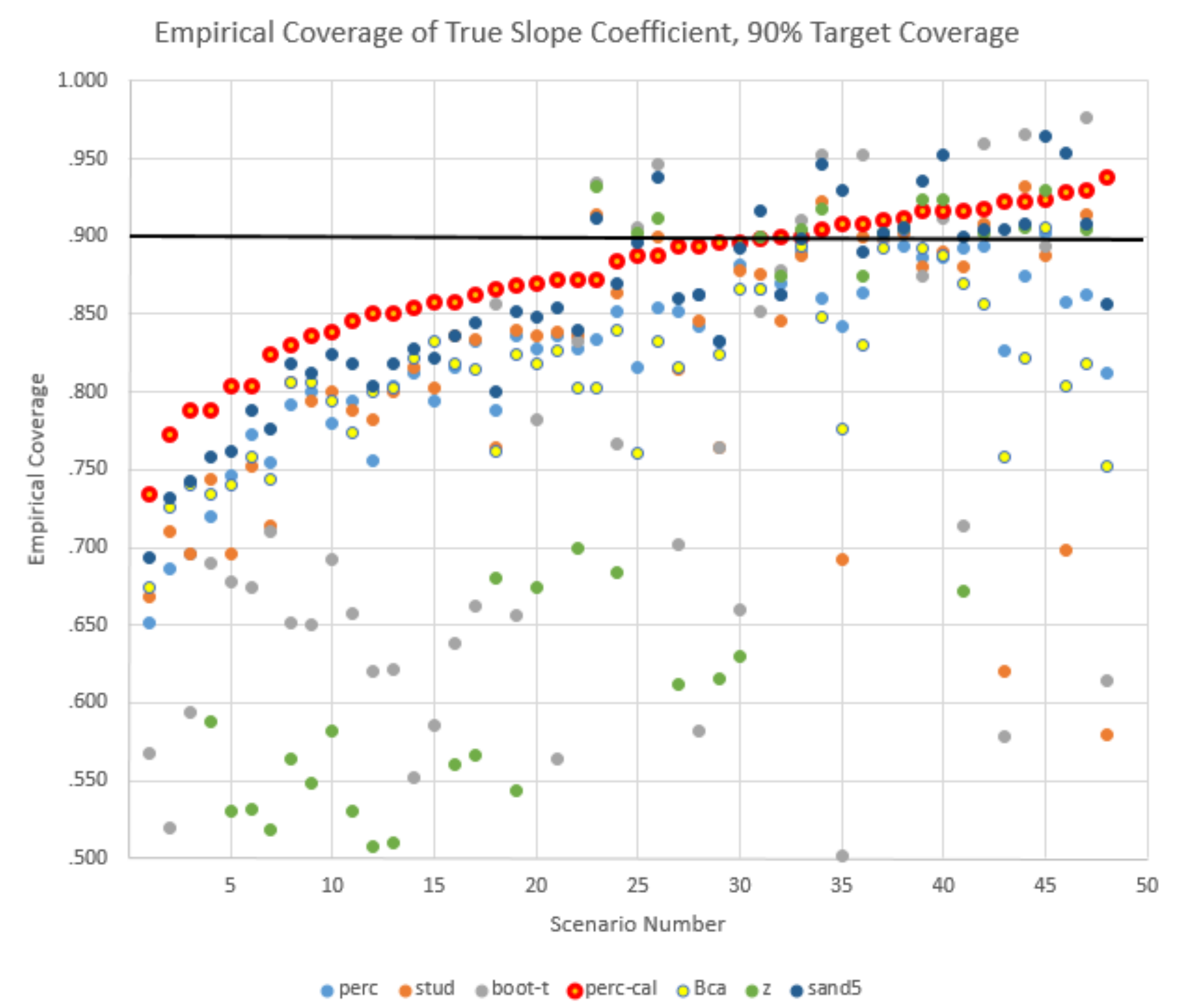}
\caption{Scatterplot of Empirical Coverage Proportion of Methods by Scenario Number -- 90\% Target Coverage}
\label{90}
\end{figure}

To more easily visualize the performance of many methods under many different scenarios, we begin with a coverage scatterplot in Figure \ref{90}. We rank sort the 48 scenarios by the empirical coverage proportion of {\tt perc-cal}, in ascending order. This ordered list determines the scenario number associated with each scenario (i.e., scenario number 1 represents the scenario in which {\tt perc-cal}'s empirical coverage was the smallest, while scenario number 48 represents the scenario in which {\tt perc-cal}'s empirical coverage was the largest). On the x-axis, we provide the scenario number. On the y-axis, we provide the empirical coverage proportion of $\beta_1$ using all methods. We exclude {\tt sand1}, {\tt sand2}, {\tt sand3}, and {\tt sand4} but include {\tt sand5}, because {\tt sand5} has a better empirical coverage proportion than the other sandwich estimators. We add a horizontal line to the graph at the desired target coverage level of 90\%.

In general, none of the methods were ``perfect'' in the sense of always providing coverage at or above the target level of coverage. All noticeably undercover in particular cases and in these cases, {\tt perc-cal}'s relative performance is generally noticeably strong. Specifically, {\tt perc-cal} achieves coverage less than target in 31 of the 48 scenarios. In 27 of those 31 below-target scenarios, {\tt perc-cal} achieved a higher empirical coverage than all other alternative methods. In the remaining 4 of the 31 below-target scenarios where {\tt perc-cal} was not the best-performing method, its absolute performance was close to target, averaging to 88.7\%, and only 1 scenario with under 88\% coverage. While {\tt perc-cal} also had the highest empirical average coverage across all scenarios, we believe this robustness to challenging scenarios is more important to practitioners, who may take comfort in the fact that when {\tt perc-cal} undercovers it almost always outperforms alternative methods, and in all other scenarios, it achieves or exceeds the desired coverage.

Across scenarios, while {\tt perc-cal} provided the most consistent empirical coverage, {\tt sand5} was itself generally superior to the other alternative methods, including {\tt BCa}, which has favorable asymptotic properties. {\tt perc-cal} achieved a mean absolute deviation from target coverage of only 3.8 percentage points, versus 5.8 and 8.9 percentage points for {\tt sand5} and {\tt BCa}, respectively. Restricting our attention to just the scenarios in which {\tt perc-cal} achieved empirical coverage below 90\%, the corresponding mean absolute deviation statistics were 5, 7.7, and 10.4, respectively. While it is perhaps no surprise that traditional methods such at the {\tt z} interval fare so poorly because they typically assume a fixed-X setting, the relative performance of these methods which do not make such assumptions is more interesting.

%

{\tt perc-cal}'s improved coverage came at the cost of modestly longer interval lengths -- across these 48 scenarios, {\tt perc-cal} had an average interval length of 1.39, at the upper end (but not the top) of other methods -- excluding the poor performance of {\tt z} and {\tt boot-t}, these other methods had interval lengths between 0.99 and 1.49, averaging to 1.16. While these other methods have interval lengths that are approximately 16\% smaller on average, it would not be acceptable to a practitioner for this shortness to come at the expense of falling below desired target coverage. Only when target coverage is achieved do considerations like average interval length become a primary concern, and 16\% longer intervals seems like a reasonable price for the ``insurance'' provided by {\tt perc-cal}.

\subsection{Real Data Example: Criminal Sentencing Dataset}

\subsubsection{Design: Real Data Example}   \label{scatterplots-dataexample}

We turn now to an example of how well {\tt perc-cal} performs in practice, on real data. In this section, we compare {\tt perc-cal} to other methods on a criminal sentencing dataset. This dataset contains information regarding criminal sentencings in a large state from January 1st 2002 to December 31st 2007. There are a total of 119,983 offenses in the dataset, stemming from a variety of crimes -- murder (899 cases), violent crime (80,402 cases), sex-related crime (7,496 cases), property-related crime (92,743 cases), firearm-related crime (15,326 cases) and drug crime (93,506 cases). An individual offense can involve multiple types of crime, and an offender's case can involve multiple charges of each type of crime.

Our modeling objective is to form marginal confidence intervals for the slope coefficients of a linear regression. The response variable of our regression is the number of days of jail time an offender must serve (log-transformed), which we predict with the following 8 covariates:
\begin{enumerate}
\item {\tt race}: Binary variable for the race of the offender (1 if white, 0 if non-white).
\item {\tt seriousness}: A numerical variable scaled to lie between 0 and 10 indicating the severity of the crime. A larger number denotes a more serious crime.
\item {\tt age}: Age of offender at the time of the offense.
\item {\tt race}: The percent of the neighborhood that is not of Caucasian ethnicity in the offender's home zip code.
\item {\tt in-state}: Binary variable for whether the offender committed the crime in his/her home state (1 if in offender’s home state, 0 otherwise).
\item {\tt juvenile}: Binary variable for whether the offender had at any point committed a crime as a juvenile (1 if yes, 0 otherwise). 18\% of all offenses involved offenders who had committed a crime as a child.
\item {\tt prior-jaildays}: Number of days of jail time the offender had previously served.
\item {\tt age-firstcrime}: The age of the offender when the offender was charged with his/her first crime as an adult.
\end{enumerate}

This is truly a random X setting because the predictors themselves are stochastic, coming to us from an unknown distribution. $\vX$ is stochastic, the relationship between $\vX$ and $Y$ is unknown and possible non-linear, and error may have heteroskedastic variance. These results are not meant to be a complete study of the issue, but rather are presented to illustrate the potential of our methodology.

We first run a linear regression upon the full dataset containing all 119,983 offenses. We treat the coefficients as if this is a population-level regression. We then proceed as if we do not have the full dataset and instead only have the ability to observe random subsets of the dataset of size 500 -- large, but not so large that all coefficient estimates are over-powered. We study the empirical coverage performance of confidence intervals formed using the methods in the simulation exercise over repeated realizations which are obtained through random subsamples of size 500.

\subsubsection{Results: Real Data Example}   \label{results-dataexample}

Linear regression across the full dataset has an $R^2$ of 16.9\% with 6 of the 8 predictors coming up as significant. We then take repeated random subsamples of size 500 from this population of offenses and treat these subsamples as if they were the observed dataset. Presupposing that each crime represents an $iid$ draw, this framework allows us to compare and contrast the empirical coverage performance of confidence interval methods.

\begin{figure}[ht!]
\centering
\includegraphics[width=0.98\textwidth]{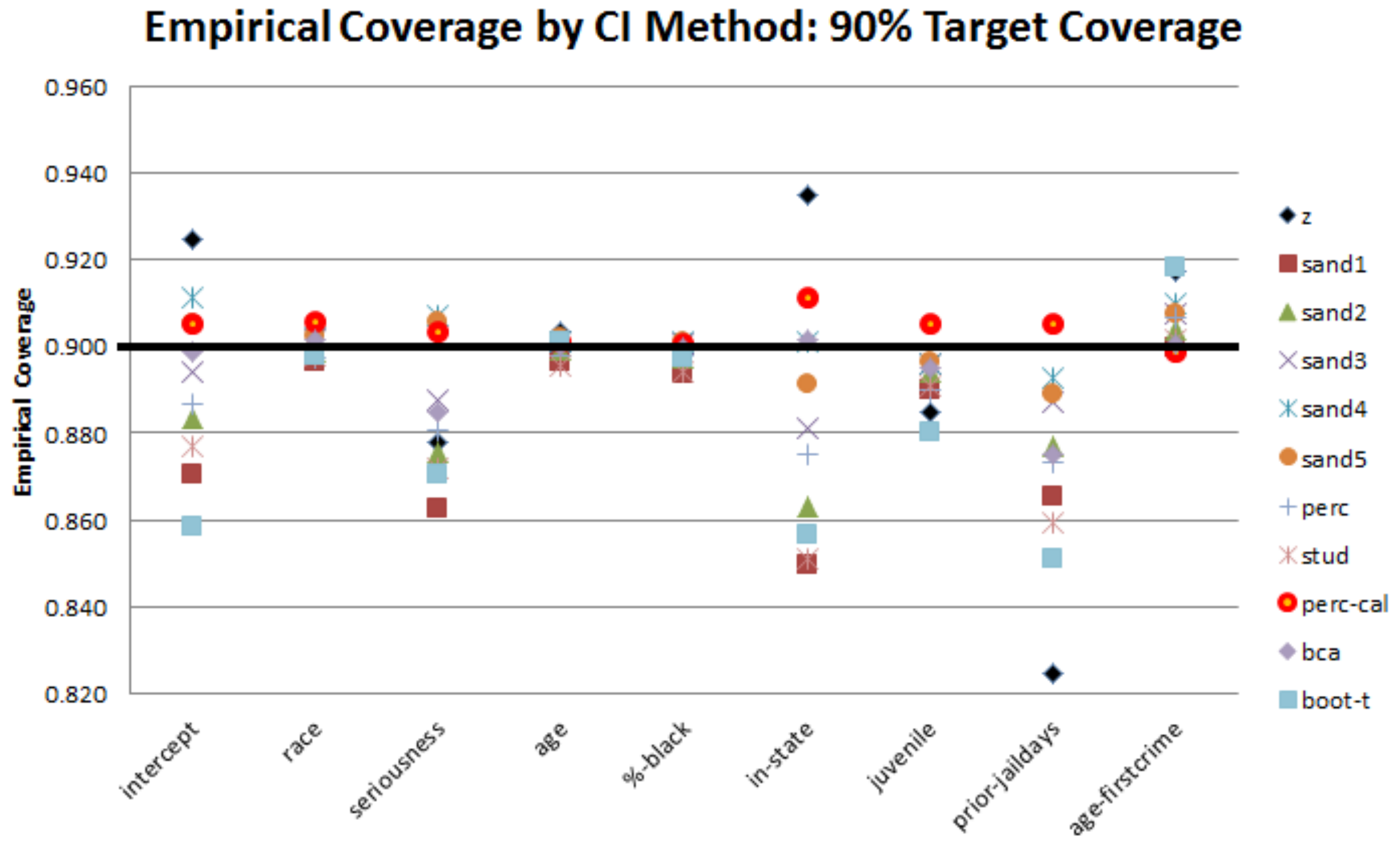}
\caption{Scatterplot of coverage proportion of methods -- 90\% Target Coverage}
\label{90-realdata}
\end{figure}

In Figure \ref{90-realdata}, we present the empirical coverage for each of our predictors when we form 90\% confidence intervals. The y-axis of the plot below represents the empirical coverage over 10,000 realizations for each of the methods in question (i.e., 90\% empirical coverage for a particular method implies that 9,000 of the 10,000 realizations had confidence intervals for that method which contained the true but unknown population-level parameters). Along the x-axis, we have the predictors listed above. We include a bold horizontal line at the target level of empirical coverage of 90\%. The standard error associated with the coverages presented below average to .002 across scenarios, predictors and methods.

There are a number of inferences that we can draw from the above chart:
\begin{itemize}
\item All methods generally perform as expected, with empirical coverage proportions generally falling between 85\% and 92\%.
\item {\tt perc-cal} is the only method that consistently achieves empirical coverage over 90\%. All other methods, including {\tt sand5}, where unable to do so.
\item {\tt prior-jaildays} appears to be the predictor with the most disappointing empirical coverage. All methods except for perc-cal do not achieve 90\% empirical coverage. The average empirical coverage of {\tt prior-jaildays} for all non-{\tt perc-cal} methods was 87.0\%.
\item There is also considerable disparity in the ability of various methods to cover the coefficients associated with the intercept term and the {\tt in-state} covariate. Although the {\tt BCa} method has near-90\% empirical coverage of the {\tt in-state} super-population coefficient, its coverage is less satisfactory for the {\tt seriousness} and {\tt prior-jaildays} covariates.
\end{itemize}

\begin{figure}[ht!]
\centering
\includegraphics[width=0.98\textwidth]{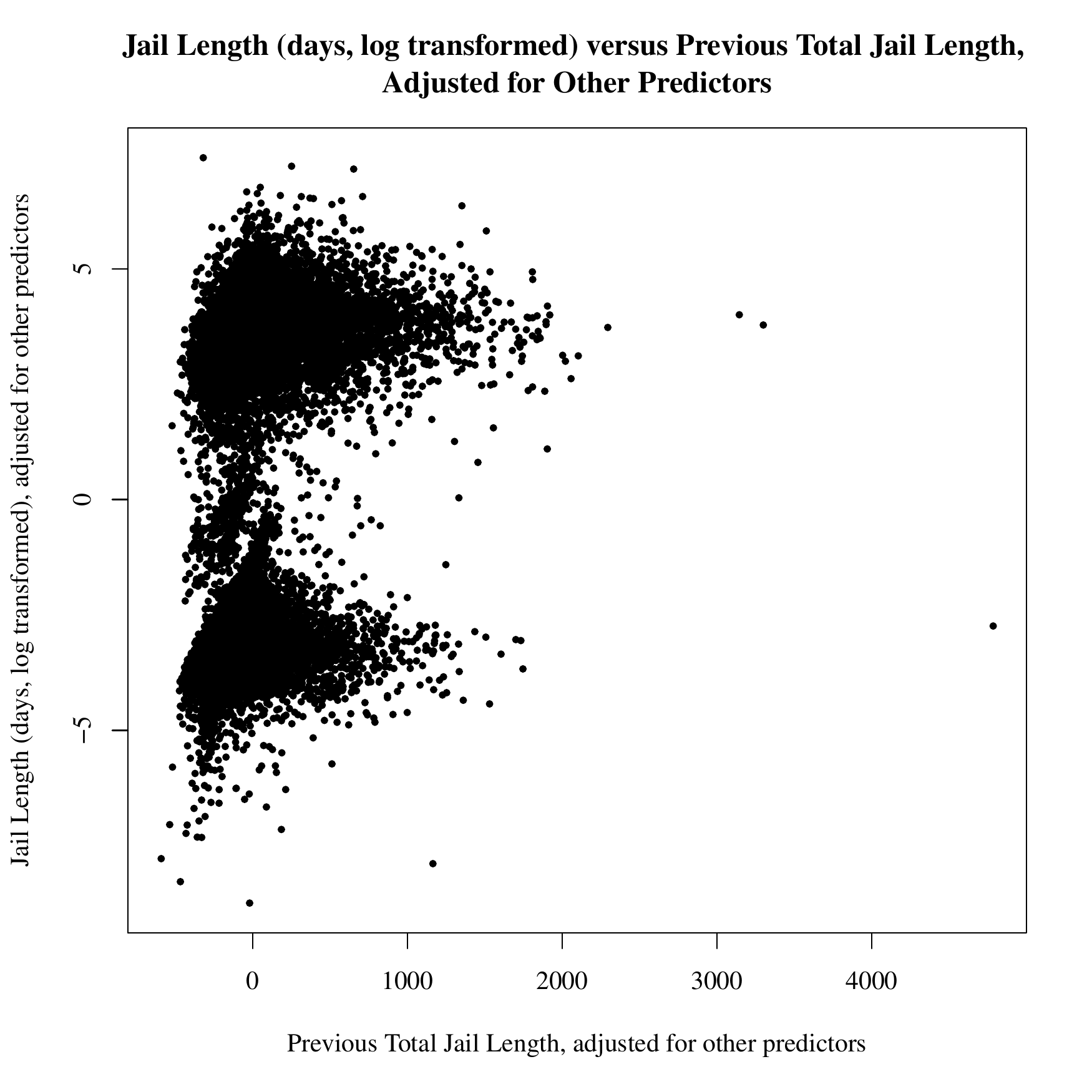}
\caption{Jail Length (log transformed) versus Previous Total Jail Length, Adjusted for Other Predictors}
\label{prior-jaildays-realdata}
\end{figure}

When we plot the relationship of jail length (log transformed) against prior total jail length in Figure \ref{prior-jaildays-realdata}, adjusted for all of the other covariates in the super-population, we see an almost bi-modal relationship.

It is clear from the plot in Figure \ref{prior-jaildays-realdata} that the highly misspecified relationship between $Y$ and $\vX$ is likely to be driving the large disparity (and general deterioration) in coverage performance across the various non-{\tt perc-cal} confidence interval methods. Overall, these results support the notion that {\tt perc-cal} is a good all-purpose confidence interval method, and that all other methods, while performing well for some of the covariates, do not perform well for all of the covariates as was the case for {\tt perc-cal}. The results assuming target coverage of 95\% are qualitatively the same as the results presented above.

\section{Discussion and Concluding Remarks}  \label{disc-concl}

If {\tt perc-cal} performs so well relative to alternative more popular CI methods, why is it not used more in practice?  We believe the use of double bootstrap methods in general have not been widely adopted primarily because of their computational cost. Although it is true that double bootstrap methods in general and {\tt perc-cal} in particular require more computation, the computational burden of these procedures is far less problematic than in the past because of current computational advances. For example, the rise of grid computing has greatly facilitated parallel computation. Because {\tt perc-cal} is trivially parallelizable, it is relatively straightforward to compute all second-level bootstrap calculations in parallel, allowing researchers to compute {\tt perc-cal} at a computational ``cost'' that is on the order of a single bootstrap. Furthermore, the perceived computational cost of double bootstrap methods may be inflated due to the inefficiency with which the calculations are carried out in popular statistical programming languages, most notably {\tt R} -- the very same calculations are orders of magnitude faster in lower level languages, such as {\tt C++}. The rising popularity and adoption of packages integrating {\tt R} with {\tt C++} (\cite{eddelbuettel2011rcpp}) can greatly reduce the cost of double bootstrap methods for practitioners performing data analysis in {\tt R} who do not know {\tt C++}. In the spirit of this, the {\tt R} package we have created allows users to compute {\tt perc-cal} intervals in {\tt R} efficiently using {\tt C++} code via {\tt Rcpp}. We are optimistic that the use of double bootstrap methods will only increase further as the cost of computing declines further over the next 10 years.

We have restricted our attention to equal-tailed intervals for all methods considered here. It is natural and certainly possible to extend our approach to compute the shortest {\it unequal}-tailed interval, even if other methods cannot or would not, because of the symmetry of the asymptotic distribution underlying those alternative methods. At the same time, this advantage should not be over-stated -- for example, one may be forced so far into the tails of the bootstrap distribution that a considerably larger number of first and second-level bootstrap samples are required. Because this is not the focus of our paper, we do not pursue it further here.

The asymptotic theory we developed, examined, and compared the more traditional percentile and ``bootstrap-t'' methods to their double bootstrap analogs in our ``assumption lean'' setting. We did not study the asymptotic properties of alternative confidence interval methods in our setting. Although it would be interesting to do so, there are a very wide range of methods in the literature, making systematic theoretical study impractical. 

In summary, randomness in $\vX$, non-linearity in the relationship between $Y$ and $\vX$, and heteroskedasticity ``conspire'' against classical inference in a regression setting (\cite{bujamodels}), particularly when the sample size is small. We have shown that in theory, the percentile-calibrated method {\tt perc-cal} provides very satisfactory empirical coverage -- the asymptotic rate of coverage error under mild regularity conditions for a two-sided confidence interval of the best linear approximation between $Y$ and $\vX$ is $\mathcal{O}(1/n^2)$. Furthermore, {\tt perc-cal} performs very well in practice, both in synthetic and real data settings. We believe that {\tt perc-cal} is a good general-purpose CI method and merits consideration when confidence intervals are needed in applied settings by practitioners.

\appendix


\section{Proof of Theorem~\ref{thm:asym}}   \label{proof-asympt}
The proof consists of two parts. The first part shows the existence of the Edgeworth expansion of the pivoting quantity. The second part derives the asymptotic order of the error term by using the first terms in this expansion and the Cornish-Fisher expansion, which can be regarded as the inverse of the Edgeworth expansion.

\noindent \emph{Part I: Existence of Edgeworth Expansion.}\\
Under three assumptions on the joint distribution in which part of the variables can be discrete, the validity of the Edgeworth expansion was shown in \cite{jensen1989} through a conditioning argument. In this section, we shall show the existence of the Edgeworth expansion of the pivoting quantity for the confidence intervals by checking these three assumptions.

We note first that by \eqref{eq:bbeta}, $\bbeta_{(p+1)\times 1}$ can be written as a smooth function of the moments $\E[X_jY]$'s and $\E[X_{j_1}^{k_1}X_{j_2}^{k_2}]$'s, where $1 \le j, \le p$, $1 \le j_1 <j_2 \le p$, $k_1, k_2 \ge 0,$ and $k_1+k_2 \le 2.$ Moreover, by the Central Limit Theorem, the asymptotic distribution of the least square estimate $\hbbeta$ in \eqref{eq:bbeta.hat} is given by
\begin{equation}\label{eq:clt}
\sqrt{n}\left(\hbbeta - \bbeta\right) \rightarrow \cN\left(\0, \E[\vX \vX^T]^{-1}\E[(Y-\vX^T\bbeta)^2\vX \vX^T]\E[\vX \vX^T]^{-1}\right).
\end{equation}
Starting with the general case when $p \ge 4$, we note that the $(p+1)\times 1$ vector of the asymptotic variances of each $\sqrt{n}(\hbeta_j-\beta_j)$, $1 \le j \le p$, is again a smooth function of the moments $\E[X_{j_1}^{k_1}X_{j_2}^{k_2}X_{j_3}^{k_3}X_{j_4}^{k_4}]$'s, $\E[X_{j_5}^{k_5}X_{j_6}^{k_6}X_{j_7}^{k_7}Y]$'s, and $\E[X_{j_8}^{k_8}X_{j_9}^{k_9}Y^2]$'s, where $1 \le j_1 < j_2<j_3<j_4 \le p$, $1 \le j_5 < j_6<j_7 \le p$, $1 \le j_8 < j_9 \le p$, $k_1,k_2,\ldots,k_9 \ge 0$, $k_1 + k_2+k_3+k_4 \le 4$, $k_5+k_6+k_7 \le 3$, and $k_8+k_9 \le 2.$ Note that these moments for the asymptotic mean and variance of $\hbbeta$ can all be consistently estimated by their corresponding sample moments.

We now collect all related monomials of $X_j$ and $Y$ into a random vector $\W_p$ such that
\begin{equation}
  \W_p =( \{X_{j_1}^{k_1}X_{j_2}^{k_2}X_{j_3}^{k_3}X_{j_4}^{k_4}\}, \{X_{j_5}^{k_5}X_{j_6}^{k_6}X_{j_7}^{k_7}Y\},\{X_{j_8}^{k_8}X_{j_9}^{k_9}Y^2\})^T,
\end{equation}
where $1 \le j_1 < j_2<j_3<j_4 \le p$, $1 \le j_5 < j_6<j_7 \le p$, $1 \le j_8 < j_9 \le p$, $k_1,k_2,\ldots,k_9 \ge 0$, $k_1 + k_2+k_3+k_4 \le 4$, $k_5+k_6+k_7 \le 3$, and $k_8+k_9 \le 2.$ For $p \geq 4$, the dimension of $\W_p$ is
$$d_p={p \choose 4}{8 \choose 4}+{p \choose 3}{6 \choose 3}+{p \choose 2}{4 \choose 2}={1 \over 12}p(p-1)(35p^2-135p+166).$$

For $1 \le p \le 3$, we can take a similar approach to write out $\W_p$.
\begin{enumerate}
  \item For $p=1$, $\W_1=(\{X_{1}^{k_1}\},\{X_{1}^{k_2}Y\},\{X_{1}^{k_3}Y^2\})^T$ where $0\le k_1 \le 4$, $0 \le k_2 \le 3$, and $0 \le k_3 \le 2.$ The dimension of $\W_1$ is $d_1=12.$
  \item For $p=2,$ $\W_{2}=( \{X_{1}^{k_1}X_{2}^{k_2}\},\{X_{1}^{k_3}X_{2}^{k_4}Y\},\{X_{1}^{k_5}X_{2}^{k_6}Y^2\})^T$ where $k_1 + k_2 \le 4$, $k_3+k_4 \le 3$, and $k_5+k_6 \le 2.$ The dimension of $\W_2$ is $d_2=31.$
  \item For $p=3,$ $\W_3 =( \{X_{1}^{k_1}X_{2}^{k_2}X_{3}^{k_3}\}, \{X_{1}^{k_4}X_{2}^{k_5}X_{3}^{k_6}Y\},\{X_{j_7}^{k_7}X_{j_8}^{k_8}Y^2\})^T$,
where $1 \le j_7 < j_8 \le 3$, $k_1,k_2,\ldots,k_9 \ge 0$, $k_1 + k_2+k_3 \le 4$, $k_4+k_5+k_6 \le 3$, and $k_7+k_8 \le 2.$ The dimension of $\W_3$ is $d_3=73$.
\end{enumerate}

With $\W_p$ defined, we can write $\bbeta_{(p+1)\times 1}=\bbeta(\E[\W_p])$. We can also write the vector of the asymptotic variances of $\sqrt{n}(\hbeta_j-\beta_j)$, $1 \le j \le p$, by $\bsigma^2_{(p+1)\times 1} =\bsigma^2(\E[\W_p])$. From the $n$ i.i.d. samples of $\vV=(X_1,\ldots,X_p,Y)^T$, we can form the $\W_p$'s as $\W_{p,1},\ldots,\W_{p,n}$. The estimates of $\bbeta$ and $\bsigma^2$ can thus be written as $\hbbeta = \bbeta(\bar{\W}_p)$ and $\hbsigma^2=\bsigma^2(\bar{\W}_p)$ respectively, where $\bar{\W}_p={1 \over n}\sum_{i=1}^n\W_{p,i}$.

To check the three assumptions in \cite{jensen1989}, we first denote the random vector of discrete variables in $\W_p$ by $\W_D$ with dimension $d_D$ and the random vector of the rest of the variables in $\W_p$ by $\W_C$ with dimension $d_C$. We shall use the same subscripts $D$ and $C$ for other related quantities to distinguish discrete variables from others. Note that in $\W_C$, some variables are products of discrete and continuous variables. The distributions of these product variables may not be absolutely continuous with respect to the Lebesgue measure. Nonetheless, the characteristic function of $\W_C$ still exists, and the proof in \cite{jensen1989} still holds with this relaxation. Thus, we continue to check these three conditions.

Assumptions $1(i), 1(ii), 1(iii), 3(i),$ and $3(ii)$ pertain to the cumulants of $\W_p$. Under the existence of the moment generating function of $\vV$, these assumptions are satisfied.

To check Assumptions $2(i)$ and $2(ii)$, note that the pivoting quantity for the confidence intervals for $\bbeta$ is $(\hbbeta-\bbeta)\circ \bsigma^{\circ (-1)} $, where $\circ$ denotes the Hadamard product. Let $\tilde{\W}_p=\bar{\W}_p-\E[\W_p]$ be the centered version of $\bar{W}_p$. The function $g(\cdot)$ in \cite{jensen1989} for regression can be written as
\begin{equation}
  g(\tilde{\W}_p) = (\bbeta(\tilde{\W}_p+\E[\W_p])-\bbeta(\E[\W_p]))\circ \bsigma(\E[\W_p])^{\circ (-1)}.
\end{equation}
It is easy to see that $g(\0)=\0.$ Furthermore, it can be shown that since the distribution of $\vV$ is non-degenerate, the derivatives of $g$ with respect to $\tilde{\W}_C$ exist, are continuous in a neighborhood of $\0$, and the $d_C \times (p+1)$ matrix of the derivatives has full rank.

To check Assumptions $1(iv)$ and $1(v)$, we first note that the derivatives of the characteristic functions are bounded by appropriate moments, which all exist under the assumptions of Theorem~\ref{thm:asym}. Therefore, we only need to show that the characteristic function of $\W_p$ is bounded away from $1$. To show this, we decompose this characteristic function by conditioning on $\W_D$, i.e.,
\begin{equation}
  \E[e^{i\t^T\W_p}] = \sum_{\w_D} e^{i\t_D^T\w_D}\E[e^{i\t_C^T\W_C}|\W_D=\w_D],
\end{equation}
where the dimensions of $\t$, $\t_D$, and $\t_C$ are $d_p$, $d_D$, and $d_C$ respectively. To verify Assumptions $1(iv)$ and $1(v)$, it now suffices to show that $|\E[e^{i\t_C^T\W_C}|\W_D=\w_D]|<1$ for each $\w_D$. This proof is patterned after the arguments in Section 2.4 of \cite{hall1992bootstrap} (page 65 to 67).

Denote the joint density of $\vV_C$ condition on $w_D$ by $f_{w_D}(\x_C,y): \RR^{p_C} \rightarrow \RR.$ We first approximate this density through simple functions. Given $\epsilon>0,$ let $f_1(\x_C,y)=\sum_{m=1}^M c_m I((\x_C,y) \in \S_m)$, where $c_m$'s are appropriate constants, $\S_m$'s are appropriate rectangular prisms, and $1 \le m \le M$ for some appropriate $M>0 $ such that $\int_{\RR^{p_C}} |f_{w_D}(\x_C,y)-f_1(\x_C,y)| d \x_C d y < \epsilon.$ We can then focus on showing that $ \lim_{\|\t_C\|_2 \rightarrow \infty} | \int_{\S_m} e^{i \t_C^T \w_C} d \x_C dy| \rightarrow 0$ for each $m$.

Let $\t_C=\t_C(u),$ where $u$ is an index that diverges to infinity. Through a subsequence argument, we can assume without loss of generality that for some $1 \le h \le d_C$, and with $s(h,\tilde{h}, u)= \t_{\tilde{h}}(u)/\t_h(u)$, the limit $s(h,\tilde{h})=\limsup_{u \rightarrow \infty} s(h,\tilde{h},u)$ exists for $1 \le \tilde{h} \le d_C$, and $|s(h,\tilde{h})| \le 1. $ We now can take $\s=\{s(h,\tilde{h},u)\}$ and write the real part of the integral $\int_{\S_m} e^{i \t_C^T \w_C} d \x_C dy$ as $  \int_{\S_m} \cos \{t_h \s^T \w_C\} d \x dy$. Note that each entry in $\w_C$ is a monomial of $x_j$'s and $y$. Thus, $\s^T \w_C$ is a polynomial of $x_j$'s and $y$. We can now take transformations of variables to show that $  \int_{\S_m} \cos \{t_h \s^T \w_C\} d \x dy = O(t_h^{-1})$. Similarly, the imaginary part of the integral can be shown to be $O(t_h^{-1})$. These facts entail that the magnitude of the conditional characteristic function is strictly less than 1 when $\|\t_C\|_2$ is large, which completes the proof.

\noindent \emph{Part II: The Asymptotic Accuracy of Double Bootstrap CIs}\\
With the existence of the Edgeworth expansion, we develop the asymptotic accuracy of the two-sided double-bootstrap CI for regression. In this section, we use $\theta_0$ to denote a generic $\beta_j$ and use $\hat{\theta}$ to denote the corresponding $\hat{\beta}_j$. We show here only the proof for the two-sided {\tt perc-cal} confidence intervals $\mathcal{I}_1$. The one-sided case for $\mathcal{I}_1$ is proved in a similar (and easier) manner. The techniques used in this proof are patterned after those in Section 3.11 in \cite{hall1992bootstrap} but are reorganized for readability and included so that our analysis is self-contained.

Consider the distribution of $\hat{A}(\bar{\W}^*) = {\hat{\theta}^*-\hat{\theta} \over \hat{\sigma}}$, where $\bar{\W}^*$ is the bootstrap version of $\bar{\W}_p$. Note that for any $0<\gamma<1$, the quantile estimate $\hat{v}_\gamma$ satisfies
\begin{equation}\label{eq:quant_v}
  \P(\sqrt{n} (\hat{\theta}^*-\hat{\theta})/\hat{\sigma} \le \hat{v}_{\gamma}|\W_{p,1},\ldots,\W_{p,n})=\gamma.
\end{equation}
Due to the existence of the Edgeworth expansion as described in Part I, we can write $\hat{v}_\gamma$ in the standard normal quantile $z_\gamma$ through the Cornish-Fisher expansion:
\begin{equation}
  \hat{v}_\gamma = z_\gamma + n^{-1/2}\hat{p}_1 (z_\gamma)+n^{-1} \hat{p}_2 (z_\gamma) + O_p(n^{-3/2})
\end{equation}
where $\hat{p}_1$ and $\hat{p}_2$ are polynomials whose coefficients are sample estimates of that of $p_1$ and $p_2,$ and the coefficients of $p_1$ and $p_2$ depend only on the moments of $\W_p$. Given the condition that all moments of $\vV$ exist, all of these estimates are root-$n$ consistent.

Now consider the quantile $\hat{w}_\lambda$ in the bootstrap distribution of $\hat{\theta}^*$ such that
\begin{equation}\label{eq:quant_w}
  \P(\hat{\theta}^* \le \hat{w}_\lambda | \W_{p,1},\ldots,\W_{p,n}) = \lambda.
\end{equation}

By comparing \eqref{eq:quant_w} and \eqref{eq:quant_v} with $\gamma=\lambda$, we see
\begin{equation}\label{eq:w_lambda}
  \hat{w}_\lambda=\hat{\theta}+ n^{-1/2}\hat{\sigma} \hat{v}_\lambda = \hat{\theta}+ n^{-1/2} \hat{\sigma}(z_\lambda + n^{-1/2}\hat{p}_1 (z_\lambda)+n^{-1} \hat{p}_2 (z_\lambda) + O_p(n^{-3/2}))
\end{equation}
Thus, by Proposition 3.1 in \cite{hall1992bootstrap} (page 102), we have
\begin{equation}
\begin{split}
  & \P(\theta_0 \in (-\infty, \hat{w}_\lambda))\\
 =& \P(\theta_0 \le \hat{\theta}+ n^{-1/2} \hat{\sigma}(z_\lambda + n^{-1/2}\hat{p}_1 (z_\lambda)+n^{-1} \hat{p}_2 (z_\lambda) + O_p(n^{-3/2})))\\
 =& \lambda +n^{-1/2}r_1(z_\lambda)\phi(z_\lambda) + n^{-1}r_2(z_\lambda)\phi(z_\lambda) +O(n^{-3/2})
\end{split}
\end{equation}
where $\phi$ is the density of the standard normal distribution, and $r_1$ and $r_2$ are even and odd polynomials whose coefficients can be root-$n$ consistently estimated.

Let $\xi=2(1-\alpha/2-\lambda)$ and $\lambda=1-\alpha/2+\xi/2.$ To find a proper $\lambda$ for the {\tt perc-cal} interval $\mathcal{I}_2$ is now to find $\xi$ such that
\begin{equation}\label{eq:quant_t}
  \P(\theta_0 \in (\hat{w}_{1-\lambda},\hat{w}_{\lambda})) =\P(\theta_0 \in (\hat{w}_{\alpha/2-\xi/2},\hat{w}_{1-\alpha/2+\xi/2})) = 1-\alpha.
\end{equation}
Note that the coverage probability of a two-sided CI can be written as
\begin{equation}
  \begin{split}
    & \P(\theta_0 \in  (\hat{w}_{1-\lambda},  \hat{w}_\lambda)) \\
  =& \P(\theta_0 \leq  \hat{w}_{\lambda}) - \P(\theta_0 \leq  (\hat{w}_{1-\lambda}))\\
  =& 2\lambda-1+ 2 n^{-1}r_2(z_\lambda) \phi(z_\lambda) +O(n^{-2}).\\
  =& 1-\alpha+\xi + 2n^{-1}r_2 (z_{1-\alpha/2+\xi/2})\phi(z_{1-\alpha/2+\xi/2})+O(n^{-2})
  \end{split}
\end{equation}
The cancellation of the $O(n^{-1/2})$ term due to that $-z_{1-\lambda}=z_\lambda$ and that $r_1$ is an even polynomial is crucial for the improvement in double-bootstrap. To achieve the accuracy of the coverage in Theorem~\ref{thm:asym}, we would like to choose $\xi$ such that
\begin{equation}
  \xi=-2n^{-1}r_2 (z_{1-\alpha/2+\xi/2})\phi(z_{1-\alpha/2+\xi/2})+O(n^{-2})
\end{equation}

Now consider the second-level bootstrap, in which we calibrate $\hat{\xi}$ for $\hat{\lambda}=1-\alpha/2+\hat{\xi}/2$ in the {\tt perc-cal} intervals. Through a similar argument for the first-level bootstrap, we see that the calibrated $\hat{\xi}$ satisfies that
\begin{equation}
  \hat{\xi}=-2n^{-1}\hat{r}_2 (z_{1-\alpha/2+\hat{\xi}/2})\phi(z_{1-\alpha/2+\hat{\xi}/2})+O_p(n^{-2})
\end{equation}
so that
\begin{equation}
  \hat{\xi}-\xi =O_p(n^{-3/2}).
\end{equation}
Finally, consider the coverage probability of the double-bootstrap CI $(\hat{w}_{\alpha/2-\hat{\xi}/2}, \hat{w}_{1-\alpha/2+\hat{\xi}/2}).$ Note that by \eqref{eq:w_lambda}, the Taylor expansion
\begin{equation}
  z_{\gamma+\epsilon}=z_{\gamma}+\epsilon \phi(z_{\gamma})^{-1} + O(\epsilon^2),
\end{equation}
and the derivations for (3.36) in \cite{hall1992bootstrap}, we have
\begin{equation}
  \begin{split}
    & \P(\theta_0 \in (-\infty, \hat{w}_{1-\alpha/2+\hat{\xi}/2})) \\
  =& \P(\sqrt{n}(\hat{\theta}-\theta_0)/\hat{\sigma}>-z_{1-\alpha/2+\hat{\xi}/2}-n^{-1/2}\hat{p}_1 (z_{1-\alpha/2+\hat{\xi}/2})-n^{-1}\hat{p}_2(z_{1-\alpha/2+\hat{\xi}/2}) +\ldots))\\
  =& \P(\sqrt{n}(\hat{\theta}-\theta_0)/\hat{\sigma}>-z_{1-\alpha/2+\xi/2}-n^{-1/2}\hat{p}_1(z_{1-\alpha/2+\xi/2})-{1 \over 2}(\hat{\xi}-\xi)\phi(z_{1-\alpha/2})^{-1} -\\
  &n^{-1} \hat{p}_2(z_{1-\alpha/2+\xi/2})+O_p(n^{-2}))\\
  =& \P(\sqrt{n}(\hat{\theta}-\theta_0)/\hat{\sigma}>-z_{1-\alpha/2+\xi/2}-n^{-1/2}\hat{p}_1(z_{1-\alpha/2+\xi/2})-n^{-1}\hat{p}_2(z_{1-\alpha/2+\xi/2})+\ldots+\\
  & {1 \over 2}(\hat{\xi}-\xi)\phi(z_{1-\alpha/2})^{-1}) +O(n^{-2})\\
  =& \P(\theta_0 < \hat{w}_{1-\alpha/2+\xi/2})+n^{-3/2} b z_{1-\alpha/2}\phi(z_{1-\alpha/2})+O(n^{-2})\\
  \end{split}
\end{equation}
where the constant $b$ is defined through
\begin{equation}
 \E[\sqrt{n}(\hat{\theta}-\theta_0)/\hat{\sigma} n^{3/2}(\hat{\xi}-\xi)/2]=b+O(n^{-1}).
\end{equation}
The $O(n^{-1})$ term is derived as in equation (3.35) in \cite{hall1992bootstrap} (page 100). Similarly,
\begin{equation}
  \P(\theta_0 \in (-\infty, \hat{w}_{\alpha/2-\hat{\xi}/2}))= \P(\theta_0 \in (-\infty,\hat{w}_{\alpha/2-\xi/2}))-n^{-3/2} b z_{\alpha/2}\phi(z_{\alpha/2})+O(n^{-2})
\end{equation}
Now
\begin{equation}
  \begin{split}
    & \P(\theta_0 \in (\hat{w}_{\alpha/2-\hat{\xi}/2}, \hat{w}_{1-\alpha/2+\hat{\xi}/2})) \\
  =& \P(\theta_0 \in (-\infty, \hat{w}_{1-\alpha/2+\hat{\xi}/2})) -\P(\theta_0 \in (-\infty, \hat{w}_{\alpha/2-\hat{\xi}/2}))\\
  =&\P(\theta_0 \in (-\infty,\hat{w}_{1-\alpha/2+\xi/2}))+n^{-3/2} b z_{1-\alpha/2}\phi(z_{1-\alpha/2})+O(n^{-2}) \\
  &- (\P(\theta_0 \in (-\infty,\hat{w}_{\alpha/2-\xi/2}))-n^{-3/2} b z_{\alpha/2}\phi(z_{\alpha/2})+O(n^{-2}))\\
  =&1-\alpha+O(n^{-2}),
  \end{split}
\end{equation}
which concludes our proof.

%
%
%
%
%
%
%

\bibliographystyle{Chicago}

\bibliography{Bibliography-MM-MC}

\end{document}